\documentclass[manuscript]{acmart}
\AtBeginDocument{%
  \providecommand\BibTeX{{%
    \normalfont B\kern-0.5em{\scshape i\kern-0.25em b}\kern-0.8em\TeX}}}

\setcopyright{acmcopyright}
\copyrightyear{2024}
\acmYear{2024}
\acmDOI{XXXXXXX.XXXXXXX}

\acmConference{CHI '24}{May 11--16,
  2024}{Honolulu, Hawai'i}
\acmBooktitle{Woodstock '18: ACM Symposium on Neural Gaze Detection,
 June 03--05, 2018, Woodstock, NY} 
\acmPrice{15.00}
\acmISBN{978-1-4503-XXXX-X/18/06}

\begin{document}

\title{EXPLORA: A teacher-apprentice methodology for eliciting natural child-computer interactions}

\author{Vanessa Figueiredo}
\affiliation{%
  \institution{University of British Columbia}
  \city{Vancouver}
  \country{Canada}}
\email{vanesfi@mail.ubc.ca}

\author{Catherine Ann Cameron}
\affiliation{%
  \institution{University of British Columbia, University of New Brunswick}
  \city{Vancouver}
  \country{Canada}}
\email{acameron@psych.ubc.ca}

%%\renewcommand{\shortauthors}{Trovato and Tobin, et al.}

\begin{abstract}
  Investigating child-computer interactions within their contexts is vital for designing technology that caters to children’s needs. However, determining what aspects of context are relevant for designing child-centric technology remains a challenge. We introduce EXPLORA, a multimodal, multistage online methodology comprising three pivotal stages: (\,1)\, building a teacher-apprentice relationship, (\,2)\, learning from child-teachers, and (\,3)\, assessing and reinforcing researcher-apprentice learning. Central to EXPLORA is the collection of attitudinal data through pre-observation interviews, offering researchers a deeper understanding of children’s characteristics and contexts. This informs subsequent online observations, allowing researchers to focus on frequent interactions. Furthermore, researchers can validate preliminary assumptions with children. A means-ends analysis framework aids in the systematic analysis of data, shedding light on context, agency and homework-information searching processes children employ in their activities. To illustrate EXPLORA’s capabilities, we present nine single case studies investigating Brazilian child-caregiver dyads (\,children ages 9-11)\, use of technology in homework information-searching.  
\end{abstract}

\begin{CCSXML}
<ccs2012>
<concept>
<concept_id>10003120.10003121</concept_id>
<concept_desc>Human-centered computing~Human computer interaction (HCI)</concept_desc>
<concept_significance>300</concept_significance>
</concept>
<concept>
<concept_id>10003120.10003121.10003122</concept_id>
<concept_desc>Human-centered computing~HCI design and evaluation methods</concept_desc>
<concept_significance>500</concept_significance>
</concept>
<concept>
<concept_id>10003120.10003121.10003122.10011750</concept_id>
<concept_desc>Human-centered computing~Field studies</concept_desc>
<concept_significance>300</concept_significance>
</concept>
</ccs2012>
\end{CCSXML}

\ccsdesc[300]{Human-centered computing~Human computer interaction (HCI)}
\ccsdesc[500]{Human-centered computing~HCI design and evaluation methods}
\ccsdesc[300]{Human-centered computing~Field studies}

\keywords{Methods, child-computer interaction, means-ends analysis.}

%%  \includegraphics[width=\textwidth]{sampleteaser}
%%  \caption{Seattle Mariners at Spring Training, 2010.}
%%  \Description{Enjoying the baseball game from the third-base
%%  seats. Ichiro Suzuki preparing to bat.}

%%\received{20 February 2007}
%%\received[revised]{12 March 2009}
%%\received[accepted]{5 June 2009}

\maketitle

\section{Introduction}
Children’s interactions with technology (\,i.e., devices and apps)\, have become integral to their lives, spanning various contexts, like home and school \cite{livingstone_parenting_2020, brazilian_internet_steering_committee_ict_2020}. These interactions are influenced by the unique aspects of each context, such as available infrastructure, technological access, personal characteristics, and social surroundings \cite{fidel_human_2012, suchman_human-machine_2006, vicente_cognitive_1999}. Given this interplay between children, technology, and their contexts, it is essential to capture Child-Computer Interaction (\,CCI)\, within their natural settings.

Yet, capturing these interactions within their contexts is far from straightforward. Previous efforts have highlighted the time-consuming and potentially disruptive nature of investigating technology use \textit{in situ} \cite{suchman_human-machine_2006, taylor_out_2011}. This challenge becomes even more evident in CCI studies, as children often require additional guidance to express their expectations and offer feedback \cite{bruckman_hci_2007}. While context-enriched methodologies like contextual inquiries have provided valuable insights, they come with their own set of challenges, especially when rapport-building with children is constrained by time and physical separation \cite{cumbo_exploring_2021, woodward_it_2022}.

Moreover, a longstanding debate centers around the translation of context-enriched discoveries into practical design recommendations \cite{dourish_implications_2006, suchman_human-machine_2006}. Despite the inclination for ethnographic accounts to shape design recommendations, these accounts hinge on interpretations that consider the interplay between participants' experiences and what the ethnographer experiences \cite{dourish_implications_2006, shapiro_limits_1994}. To mitigate the risk of overly simplistic \textit{conceptualization of the setting}, means-ends analyses have offered a framework to methodically investigate the interplay between contextual aspects and interaction patterns \cite{fidel_information_2004, hertzum_analysis_2002}. A means-ends analysis (\,MEA)\, focuses on the analysis of the interplay between context, cognitive processes (\,i.e., problem-solving and decision-making)\, and interactions in affecting how people achieve their goals (\,ends)\, while using technologies (\,means)\, \cite{allen_hierarchy_2017, rasmussen_taxonomy_1990}. Previous work applied MEA to identify the relationship between work functions, resources and system limitations, generating detailed interaction scenarios to inform design recommendations \cite{rauffet_designing_2015, kwon_evaluation_2007}. While the application of MEA frameworks to CCI studies remains untapped, it presents a potential avenue for delivering insights centred on frequent context-based interactions.

We introduce the EXPLORA methodology—Exploration Through Teaching and Apprenticeship—an approach that revolves around the interplay of context, interactions, transactions, and agency. Context, defined as an interconnected system involving settings, situations, and constraints \cite{vicente_cognitive_1999}, influences how children engage (\,i.e., agency)\, with technologies and the evolution of these interactions over time (\,i.e., transactions)\,.

EXPLORA facilitates frequent child-computer interactions while empowering children to contribute as experts in shaping technology solutions based on their experiences. It is a multimodal, multistage online methodology comprising three pivotal steps: (\,1)\, building a teacher-apprentice relationship, (\,2)\, learning from participant-teachers, and (\,3)\, evaluating and reinforcing researcher-apprentice learning.

Central to the EXPLORA approach is the collection of attitudinal data through pre-observation interviews, offering researchers a deeper understanding of participants’ journeys, needs, and interactions. This informs subsequent online observations, allowing researchers to focus on relevant contextual aspects. The methodology also emphasizes validating preliminary assumptions with participants, ensuring consistency with their experiences. A means-ends analysis framework aids in the systematic analysis of data, shedding light on decision-making, cognitive processes, and strategies employed by participants in tasks.

To exemplify the practicality and potential of EXPLORA, we present nine single case studies involving Brazilian child-caregiver dyads (\,children ages 9-11)\, using technology (\,i.e., devices and apps)\, during homework information searches. The first author, a native Brazilian-Portuguese speaker, facilitated access to the Brazilian contexts presented in the case studies. Our findings reflect the success of the methodology in fostering open communication and comfort among participants, especially children, by utilizing children’s terminologies. The guidance provided during the study helped participants align their experiences with research objectives, facilitating the unveiling of insights. The integration of the means-ends analysis framework further provided a comprehensive perspective on technology preferences and strategies used in homework information-searching.

Our contribution lies in providing a holistic framework that captures context-rich data while unravelling the intricacies of children’s decision-making and cognitive processing. While means-ends analysis approaches have been applied extensively in HCI studies involving adults \cite{fidel_information_2004, hertzum_analysis_2002}, we bring a novel contribution by developing a methodology focused on means-ends analysis that is aimed at children. Furthermore, this child-centric systemic approach identifies the systems in which children interact with technology, uncovering the factors influencing CCI within different contexts. Ultimately, our methodology offers a comprehensive and structured way to elicit, integrate, and contextualize attitudinal and behavioural insights from child participants, opening new avenues for research in the CCI.
\section{Background}
\subsection{Context-Enriched Methodologies}
Context-enriched methodologies refer to approaches aimed at gaining a comprehensive understanding of how users interact with devices and apps within their natural contexts. We will discuss context-enriched methodologies utilizing behavioural, attitudinal (\,i.e., self-reported)\, and a combination of behavioural-attitudinal methods. 
\subsubsection{Behavioural.} Ethnography has been a traditional behavioural-oriented context-enriched methodology in HCI \cite{rogers_research_2017, suchman_human-machine_2006}. Rooted in anthropological traditions, ethnography allows the investigation of personal characteristics, processes, and settings \cite{taylor_out_2011}. However, the unstructured nature of ethnographic observations can make it challenging to detect relevant interactions from the sheer volume of information \cite{shapiro_limits_1994}. Additionally, ethnographies tend to focus on describing observed interactions rather than providing in-depth analyses of the underlying reasons for these behaviours within the contextual landscape \cite{dourish_implications_2006, suchman_human-machine_2006}. Ethnographies demand negotiations to gain authorization for accessing the setting, along with caution to ensure that the researchers’ presence does not disrupt the users’ routines or the environment's natural dynamics \cite{bruckman_hci_2007, taylor_out_2011}.

Other behavioural methodologies, such as usability evaluations in context, also exist, but they often restrict the specific aspects of context to be emphasized. For instance, field deployments adopt an interventionist approach that influences people's daily routines by introducing new technologies to evaluate them in context \cite{siek_field_2014}. To comprehend changes in these behaviours, researchers typically need to carry out longitudinal studies.
\subsubsection{Attitudinal.} The complexities associated with conducting ethnographies have instigated the exploration of remote methodologies supported by participants’ self-reports. Asynchronous remote usability evaluations, such as the experience sampling method, entail participants sharing reports of their experiences with digital apps deployed in their natural contexts \cite{consolvo_using_2003, larson_experience_1983, stone_ecological_1999}. Other attitudinal methodologies, such as diary studies \cite{wisniewski_dear_2016}, photo diaries \cite{barriage_mobile_2020}, cultural probes \cite{hong_using_2020, riekhoff_sampling_2008}, and technology probes \cite{cagiltay_designing_2023, hutchinson_technology_2003} encourage children to engage in a variety of expressive modes—writing, photography, drawing, and recording—to document their experiences in natural contexts. While these approaches grant researchers access to children’s contexts, behaviours, and interactions, the gap between what is reported and the nuanced context of the interaction can sometimes limit the depth of insight researchers can gain \cite{barriage_mobile_2020}.
\subsubsection{Behavioural-Attitudinal.} We refer to contextual inquiries as those methodologies that utilize multi-method study design that combine behavioural and attitudinal methods, such as natural observations, interviews, diary studies and workshops. While contextual inquiry precedes a contextual design stage, where design teams identify frequent interactions \cite{holtzblatt_contextual_2014}, some contextual inquiry studies emphasize contextual inquiry rather than progressing to the subsequent contextual design phase \cite{kaplan_reading_2006, wallbaum_towards_2017}.

An interesting aspect of contextual inquiry relies on a \textit{teacher-apprentice} relationship formed between researchers and participants \cite{holtzblatt_contextual_2014}. The \textit{teacher-apprentice} relationship establishes how participants can contribute to research by encouraging them to demonstrate their interactions with technology while providing insights into their problem-solving and decision-making processes. While the \textit{teacher-apprentice} relationship has gained popularity with contextual inquiries, the \textit{teach-me-back} approach, which originated in the 1970s, pioneered in prompting participants to externalize their cognitive processes (\,i.e., problem-solving and decision-making)\, by explaining their actions \cite{pask_learning_1972}. Recent iterations involve participants drawing their ideas to support the articulation of their choices and strategies \cite{van_der_veer_mental_1994}. Marhan and colleagues \cite{marhan_review_2012} argued that drawings would encourage children to explain their ideas about technology. While a drawing approach may work in participatory design approaches, children may require a more explicit prompt to demonstrate what they do to achieve their tasks in context and externalize cognitive processes to explain their behaviours \cite{bruckman_hci_2007}. For instance, children produced detailed narratives when receiving guidance and prompts to recall past experiences \cite{cameron_elicitation_1996}.

Among the array of context-enriched methodologies, the potency of a guided multimodal approach, as exemplified by contextual inquiries, provides a path to eliciting authentic insights from children regarding their natural interactions and behaviours involving the use of technology.
\subsection{Contextual inquiries}
While traditional contextual inquiries involve observations and interviews, CCI studies have applied other methods to accommodate children’s communication skills, study motivations and attention span \cite{marhan_review_2012, ferron_walk_2019}. The multimodal aspect of contextual inquiries has helped researchers obtain rich accounts of children’s interactions with digital apps in context. For example, these accounts encompass finding nuances between how caregivers and children perceive and interact with technology \cite{hubbard_voice_2021}, and identifying moments in which children, educators and caregivers face challenges to achieve their tasks in educational apps \cite{marcu_breakdowns_2019}. In ecological-oriented contextual inquiries, the analysis of proximal (\,e.g., family, access to devices)\, and distal (\,e.g., school policies)\, contextual factors revealed that children’s interactions with apps do not depend exclusively on the constraints posed by these digital apps \cite{kalinowski_ecological_2021}.

Method sequence in contextual inquiries can impact the understanding of relationships between interactions and context, especially in identifying what aspects of context should be further analyzed. When interviews were conducted before observations, researchers identified what technologies were utilized, children’s perceptions, frequent activities and common terminologies. For example, Nansen and colleagues \cite{nansen_children_2012} identified the communication modes and literacies children were inclined to use prior observations, which might have helped children discuss their experiences with apps. When observations were conducted first, researchers followed up with interviews and workshops focused on clarifying interactions and the rationale behind them and validating contextual relationships \cite{ferron_walk_2019, gillen_day_2018}.

The method sequence becomes more crucial in online contextual inquiries, as researchers and participants are not co-located. As rapport building is limited due to physical and time constraints, children may require incentives, such as detailed prompts providing a step-by-step of how they can contribute to the study \cite{lee_unboxing_2022}. Cumbo and colleagues \cite{cumbo_exploring_2021} interviewed children and their parents first to outline the homeschooling context before the photo-diary step. Because the study did not involve observations, researchers relied on the data captured in the first interviews to encourage children to discuss the pictures taken during the photo-diary stage.
\subsection{Conceptual Framework}
We adopt a quasi-ecological perspective in our methodology to explore child-computer interactions in context. Ecological methodologies are designed to transcend the confines of isolated observations, delving into the dynamic and interconnected systems where children live. Our methodology relies upon three conceptual pillars: context, interactions/transactions and agency detailed in the following sub-sections.
\subsubsection{Context.} Depending on the epistemological stance researchers take, context can be the physical space, a social construct, an event or a system \cite{dervin_given_2003}. We consider context as an interconnected system encompassing settings, situations, and constraints that affect how children interact and transact their behaviours when utilizing devices and apps.
\subsubsection{Interactions/ Transactions.} Interactions encompass actions, inputs, and outputs involved in users’ engagement with technology (\,devices and apps)\,. While interactions have been extensively explored in HCI \cite{bruckman_hci_2007}, the transformations in users’ perception and understanding of technology are often embedded in the analysis of interactions. An interaction that potentiates change in either of the interactants is seen as a transaction.  Either party can be changed to transform an interaction into a transaction. These transformative interactions—transactions— lead to developmental changes, particularly in children \cite{fogel_what_2009}.
\subsubsection{Agency.} Agency is the capacity of a person to exercise and manifest their actions based on internal (e.g., cognitive processes)\, and external conditions (\,e.g., other people, availability of resources, environment)\, \cite{bandura_toward_2006}. Thus, agency plays a crucial role in shaping children's choices on how to engage with devices and apps, as well as in their problem-solving and decision-making processes related to such interactions.
\subsubsection{Means-Ends Analysis}. A long-time discussion in HCI refers to translating context-enriched findings into design recommendations \cite{barkhuus_mice_2007}. Rather than exploring the \textit{how} and \textit{why}, Shapiro \cite{shapiro_limits_1994} argued that most ethnographies result in descriptive accounts of observable behaviours. Despite the tendency for ethnographic reports to translate into design recommendations, these reports rely on interpretations of how participants’ experiences are interpreted through the ethnographer’s understanding of that experience \cite{dourish_implications_2006}. To counter the \text{conceptualization of the setting}, Dourish \cite{dourish_implications_2006} advocated for the use of frameworks that theorize and organize research findings.   
Means-ends analyses are well-established approaches in the field of human factors and ergonomics, offering a systemic analysis that can have valuable applications in HCI \cite{fidel_information_2004, hertzum_analysis_2002}. While multiple variations of means-ends analysis exist \cite{dutta_engineering_2013, young_using_1990}, we focus on the Cognitive Work Analysis (\,CWA)\, framework \cite{fidel_information_2004, rasmussen_taxonomy_1990, vicente_cognitive_1999}. The CWA means-ends analysis involves decomposing complex tasks or activities into distinct steps (\,\textit{means})\, required to achieve specific goals (\,\textit{ends})\,. This analysis aims to understand the underlying cognitive processes and strategies that users employ to accomplish tasks while considering the broader context in which these interactions occur. In the MEA, processes encompass the dynamic interactions, transactions, and cognitive processes involved when an individual accomplishes a task. Therefore, the MEA delves deeper than superficial interactions, revealing the cognitive workload, decision-making, and strategies users employ while navigating through tasks.  
While the CWA means-ends analysis has not been applied in child-computer interaction studies, it provides a possible avenue to deliver actionable insights focusing on frequent interactions in context. CWA means-ends analysis enables researchers to systematically map out the interplay between contextual elements and interaction patterns, shedding light on how children adapt their behaviours based on different situations.
\section{EXPLORA}
The Exploration through Teaching and Apprenticeship methodology—EXPLORA—focuses on interactions in the context of technology use. This methodology is grounded in experiential learning principles, where participants (\,such as children and caregivers)\, assume the role of teachers, guiding researchers (\,the apprentices)\, through their problem-solving and decision-making processes during computer interactions. The core objective of the EXPLORA methodology is to capture frequent, fine-grained child-computer interactions, while also empowering children as experts in shaping technology solutions based on their own experiences.
The EXPLORA methodology encompasses three stages (\,1)\, Build a teacher-apprentice relationship, (\,2)\, Learn from the participant-teacher, and (\,3)\, Assess and reinforce the researcher-apprentice's learning.

In the \textit{Building a Teacher-Relationship} stage, participants gain an understanding of their contribution to the study and reflect on their experiences before data collection starts. While the teacher-apprentice relationship develops over the study's duration, it is crucial to introduce and emphasize the metaphor during our initial interaction with participants. Drawing from the teacher-apprentice metaphor, we position ourselves as learners, acknowledging our limited familiarity with participants' technology use. Moreover, we highlight the participant's role as the expert in their own life, underlining our interest in learning from them. In this initial stage, we outline the study's objectives, stress the significance of child-caregiver dyads' insights, provide a brief overview of the data collection process, and collect written consent and assent.

During the \textit{Learning from the Participant-Teacher} stage, we encourage participants to openly discuss and share their experiences. Whether participants are being interviewed, observed or recalling past occurrences, we concentrate on identifying instances that hold frequent value. When such instances arise, we pose a guiding question — \textit{Can you teach me [this]?} — to participants. This question prompts participants to illustrate and clarify their task performance and technology usage, offering valuable insights into the constraints influencing their interactions. The guiding question offers minimal direction and helps participants elucidate their technology interactions within their contexts. During observations, participants explain the natural task scheduled for the session. Subsequently, we utilize the guiding question to assist participants in conveying and showcasing their task execution. This question — \textit{Can you teach me [this]?} — remains pivotal as it functions as the catalyst propelling participants to educate researchers about their experiences and technological engagements.

The \textit{Reinforcing Learning} stage involves revisiting our preliminary assumptions and inviting participants to evaluate our acquired knowledge. We review our notes and video recordings, focusing on frequent instances that embody child-computer interactions within the context. We then interpret these instances based on our gathered insights. Following this, we present our preliminary assumptions to participants and ask their assessment of accuracy. This approach not only diminishes bias but also facilitates a comprehensive review, aligning it more closely with natural child-computer interactions.
\subsection{Multimodal, Multistage Approach}
Our study design followed a sequence informed by data from previous sessions, enabling the development of protocols and formulation of relevant questions for subsequent sessions, including observations, events-based sessions, and reflective interviews. This sequential approach helped identify terminologies, preferences, and interactions, guiding our focus in subsequent sessions. The following sections describe the methods in the sequence order designed for our study.
\subsubsection{Preliminary Session.} The goals of this session were to (\,1)\, introduce the study and how participants would be involved; (\,2)\, address participants’ concerns; (\,3)\, conduct a 5-minute demonstration of what the observation would look like; (\,4)\, schedule the first study session; (\,5)\, collect consent and assent agreements.
\subsubsection{Interviews.} We initiated the interviews as a conversation centred around the practices of searching for homework information. To maintain an unbiased perspective, we began by asking about the child's day. Subsequently, we prompted the child with  \textit{Can you teach me [this]?} to elicit their insights when they mentioned any activities related to information searching using devices and apps. We applied a similar approach when interviewing the caregiver.
\subsubsection{Observations.} To select the most fitting natural task that would spotlight homework-related information searching involving technology use, we requested caregivers to be mindful of instances when their children were assigned a homework task likely to require the use of technology for information searching. Caregivers were then requested to reach out to us to schedule a session, during which the child-caregiver dyad would engage in natural homework information-searching processes. The child-caregiver dyad had time to address any concerns they had before the beginning of the video recording. Our focus was on capturing the child-caregiver's interactions/transactions via the webcam and tracking screen actions using the screen-sharing functionality offered by video conferencing software. As the video recording started, we initiated by asking the child: \textit{Can you teach me [this]?} This prompt aimed to encourage dyads, especially children, to demonstrate regular homework information-searching processes without feeling evaluated. While the prompt could influence the dyads, our goal was to reduce bias by emphasizing their expertise and eliciting open and uninhibited homework information-searching processes.
\subsubsection{Events-based sessions.} While the interviews focused on eliciting the child-caregiver dyad's perspectives on frequent interactions and transactions involving homework information-searching, the events-based sessions focused on the child's report or re-enactment of recent homework information-searching interactions and transactions involving the use of apps and devices. The events-based sessions presented additional opportunities to witness child-caregiver dyads engaging in natural homework-related tasks. After the child reported recent past homework information-searching events, we asked the prompt question: \textit{Can you teach me [this]?} to elicit what the child-caregiver dyad had done.
\subsubsection{Reflective Interview.} We included a reflective interview session to clarify whether our preliminary understanding accurately reflected the online contexts under study. We analyzed notes taken during the interviews, observations and critical incident sessions. The selected notes highlighted interactions and aspects affecting the homework information searching process of the child-caregiver dyads in the study.

Our multimodal, multistage approach minimized the time and effort required from participants during each session. Furthermore, the multistage study design allowed us to capture ongoing oral consent and assent before each session started, and regular check-ins to address participants' concerns. Other than including detailed research protocols for each stage, including problem-solving strategies (\,e.g., if the participant does not answer question A, the researcher will ask question B and its rationale)\,, our research proposal for the university’s Institutional Ethics Review Board addressed ethical, privacy and data management concerns.
\subsection{Data Analysis}
Following the completion of each study session, we composed memos detailing the preferred terminologies used by the child-caregiver pair, contextual aspects pertinent to their homework information-searching contexts, and our initial assumptions. To gain familiarity with the data, we reviewed the video recordings within a two-week window from the data collection date, capitalizing on the freshness of our recollection. Subsequently, we referred to our memos to determine which data should undergo transcription and how it should be prepared for this process. In our case, there was an additional requirement to translate the data from Brazilian Portuguese to English.

The data was analyzed using a thematic analysis approach. We applied a deductive thematic analysis focusing on the hierarchical levels in the MEA framework. While a deductive thematic analysis may yield a framed description of the data, the goal of applying a deductive thematic analysis first was to code the data to facilitate the MEA. Then, we applied an inductive thematic analysis to understand the interplay between context, interactions, transactions and agency concerning the means (\,i.e., devices and apps)\, children utilized in their homework information-searching processes. For the inductive thematic analysis, we watched the videos, took notes and read the translated transcriptions, listing potential themes. To guide our inductive thematic analysis, we prioritized developing codes that provided a connection between the three pillars (\,context, interactions/transactions, and agency)\,. After individually coding the data, we compared our themes to consolidate the theme book, theme description and representative quotes.
\section{EXPLORA Case Studies}
\subsection{Methods}
We first applied the EXPLORA methodology in a series of single case studies to understand how children (\,ages 9-11)\, utilized digital technologies to accomplish schoolwork in Brazilian contexts. The study spanned seven weeks and involved nine Brazilian child-caregiver dyads (\,table \ref{tab:demographics})\,. The university's institutional review board approved our research proposal (\,Ethics Certificate H20-03568)\,. 

\begin{table}[h!]
    \centering
    \begin{tabular}{ccccc}
        \textbf{Pseudonyms} & \textbf{School} &  \textbf{Child's Age}&  \textbf{Child's gender}& \textbf{Caregiver's Educational Level}\\
        \hline
        \vtop{\hbox{\strut Iris (\,CH)\,}\hbox{\strut Clara (\,CA)\,}} & Gamma & 9  & F & Bachelor degree\\
        \hline
        \vtop{\hbox{\strut George (\,CH)\,}\hbox{\strut Audrey (\,CA)\,}} & Gamma & 9  & M & Secondary\\
        \hline
        \vtop{\hbox{\strut Stephanie (\,CH)\,}\hbox{\strut Diana (\,CA)\,}} & Omega & 11 & F & Elementary\\
        \hline
        \vtop{\hbox{\strut Maria (\,CH)\,}\hbox{\strut Rita (\,CA)\,}} & Gamma & 10 & F & PhD\\
        \hline
        \vtop{\hbox{\strut Russell (\,CH)\,}\hbox{\strut Alice (\,CA)\,}} & Kappa & 9  & M & Bachelor degree\\
        \hline
        \vtop{\hbox{\strut Philip (\,CH)\,}\hbox{\strut Brenda (\,CA)\,}} & Gamma & 9  & M & Bachelor degree\\
        \hline
        \vtop{\hbox{\strut Noah (\,CH)\,}\hbox{\strut Alana (\,CA)\,}} & Zeta  & 9  & M & Elementary\\
        \hline
        \vtop{\hbox{\strut John (\,CH)\,}\hbox{\strut Joanna (\,CA)\,}} & Iota  & 9  & M & Bachelor degree\\
        \hline
        \vtop{\hbox{\strut Levi (\,CH)\,}\hbox{\strut Patricia (\,CA)\,}} & Theta & 10 & M & Secondary\\
        \hline
    \end{tabular}
    \caption{Child-Caregiver Demographics. CH = child, CA = caregiver}
    \Description{The table illustrates the demographics of the nine child-caregiver dyads in our study.}
    \label{tab:demographics}
\end{table}

We recruited participants with the assistance of a local Brazilian mediator within our research network. One month before the study started, we met with this local advocate, who is a mother to a 9-year-old Grade 4 student. During our initial meeting, we covered the study's purpose, its stages, how participants would be involved, privacy concerns, and study compensation. Following this, the mediator distributed our study invitation through a WhatsApp group chat. This group consisted of caregivers (\,e.g., mothers, fathers and grandparents)\, whose children attended the same Grade 4 class as the mediator's daughter. Initially, we received responses from only four caregivers. To expand our participant pool, we employed a "snowball" sampling approach, asking these four caregivers to share the invitation with others. Through this strategy, an additional five caregivers volunteered to participate. Although we did not specify the caregiver role (\,mother, father, or grandparent)\,, only mothers responded to our recruitment letter. As compensation for their participation, each dyad received a \$10 gift card per completed session, totalling \$70.

We aimed to gain insights into how children employed devices (\,such as smartphones, tablets, and laptops)\, and apps (\,including conversational agents, social media, and search engines)\, to fulfill their homework information-searching processes. Specifically, we sought to understand the problem-solving and decision-making processes behind their choice of devices or apps and the underlying rationale. Our focus was on observing children as they completed or discussed natural homework tasks, to understand the impact of contextual factors on their problem-solving and decision-making.

To assess the efficacy of EXPLORA in these case studies, we conducted a secondary data analysis. We revisited the videos and reviewed the transcripts to identify instances where EXPLORA helped elicit instances concerning \textit{what}, \textit{how}, and \textit{why} involving homework information-searching processes, insights for design recommendation and the benefits and challenges of this methodology.

\subsection{Findings}
The subsequent sub-sections will delve into the four themes emerging from the secondary analysis: (\,1)\, Method sequence, (\,2)\, Dyads as teachers, (\,3)\, Contextual relationships, and (\,4)\, Design recommendations.
\subsubsection{Method Sequence}
Our multimodal approach enabled us to capture various perspectives on how dyads approached homework information-searching processes. Additionally, it aided in understanding terminologies, preferences, and constraints in the analyzed contexts, as we strategically determined the sequence of methods during data collection.

Starting with interviews proved beneficial as it allowed us to identify how dyads described their homework information-searching processes and the reasoning behind their choices among different processes.

\begin{quote}
    Maria: I use my phone and my mom’s laptop all the time and at the same time to do my homework.\par
    R: Can you teach me how you use your phone and your mom’s laptop at the same time to complete your homework?\par
    Maria: If I need to do a project, I look there on TikTok, then go to the laptop because I can watch other videos or find texts. Here’s what I do: I get the video title from TikTok and go to the laptop. I type the video title on Google and check the links.\par
    \textbf{Interview}
\end{quote}

We asked Maria to discuss the devices and apps she used when searching for homework information. This preliminary discussion was essential to set the stage for the \textit{Can you teach me [\,this]\,} prompt and what would follow after asking the child to teach us. Maria's insights provided a comprehensive view of how she decides on devices or apps and the specific features aiding her homework information-searching strategies. Based on these insights, we identified instances for further exploration, remaining open to spontaneous developments during subsequent observations and events-based sessions. For instance, Maria utilized TikTok during the observation session to search for a song for a school project, revealing her tendency to switch between devices and apps for specific tasks.
\subsubsection{Dyads as Teachers}
We established a \textit{teacher-apprentice} relationship with the child-caregiver dyads across the three stages of our study. The process of building this relationship started before data collection in the preliminary session. We found that most dyads (\,7)\, grew accustomed to taking on the role of a teacher, becoming more comfortable with reporting and demonstrating their homework information-searching processes over time.

\begin{quote}
    R: Stephanie, would you like to be my teacher?\par
    Stephanie: Yes!\par
    Diana: She found her place. She loves teaching everyone, like her dad, her grandparents, her sister.\par
    \textbf{Stephanie (\,child)\,- Diana (\,caregiver)\,, Preliminary Session}
\end{quote}

Stephanie showed enthusiasm about taking on the role of a teacher in our study. She, along with two other children, prepared \textit{"lesson plans"} to teach us during our study sessions, noting their activities in the days leading up to the scheduled sessions.

Nevertheless, it was not always natural for children to take charge, particularly when interacting with unfamiliar adults.

\begin{quote}
    R: Noah, would you like to be my teacher?\par
    Noah: I don’t know. I don’t know how to teach grown-ups.\par
    Alana: Come on, don’t be shy.\par
    Noah: Why do you want to know about my homework\par
    R: It’s been a long time since I was at school. So, I need to know what kids are doing now because this is part of my homework.\par
    \textbf{Noah (\,child)\,- Alana (\,caregiver)\,, Preliminary Session}
\end{quote}

Noah appeared uncertain about assuming the role of a teacher in this study. It was essential to ensure his comfort with the role by explaining how his contributions would help us with our \textit{homework}. Noah displayed shyness during the interview session and refrained from answering most questions about his homework information-searching routine. However, as we learned that Noah's favourite hobby was playing games during the preliminary session and that he frequently searched for game-related information, we opted to inquire about his approach to finding game information. Not only did he teach us his game information-searching processes, but this also instilled confidence in him to share insights into his broader homework information-searching practices.

Defining teacher(\,s)\, and apprentice(\,s)\, roles was also fundamental for eliciting and capturing natural homework information-searching processes, particularly problem-solving strategies and agency.

\begin{quote}
    R: Before we start, I want to say that you’re my teacher, John. I know nothing about how you do your homework, ok?\par
    John: Ok. Do you want me to read the homework description, so you know what this homework is about?\par
    R: Yes. Teach me however you think it’s the best.\par
    John: I'll go and check on WhatsApp if my friends posted something.\par
    [\,John opens WhatsApp on his phone]\,\par
    John: Nothing.\par
    [\,John gets his tablet and activates Google Assistant]\,\par
    John: Hey Google, what is it called the plain areas in Brazil?\par
    John: Hey R, I chose Google Assistant because I was only checking how I'd type keywords on Google (\,search)\,.\par
   \textbf{Observation}
\end{quote}

After emphasizing the teacher and apprentice roles at the beginning of the session, John showed us in detail how he searched for information to complete his homework, assuming the role of a teacher. Although John's decision to delve into the specifics of his actions might be seen as reducing the spontaneity of his homework information-searching process, it afforded us insight into his reasoning for choosing Google Assistant first and then typing queries into Google Search.

During the events-based sessions, the dyads discussed various events involving homework information searching, including those in which the dyads considered that they were not successful in achieving the expected results. 

\begin{quote}
    Iris: Look, I have a list with all the things I did last week. I can choose something from the list to teach you.\par
    R: Ok. What do you want to teach me today?\par
    Iris: I want to teach you how I searched for the answer to my cardinal directions homework.\par
    \textbf {Events-based session \#2}
\end{quote}

Iris chose to recreate her homework information-searching processes, noting, that \textit{"this homework was fun because I searched for videos and pictures"}. In another instance during the same session, Iris replicated a homework information-searching process for a task requiring her to list the counties in her city. While re-enacting, she mentioned, \textit{"had to ask my dad because I couldn't find anything online but now I know how to search for counties online"} and demonstrated what she had learned from her father. Encouraging the dyads, especially the child, to teach us about challenging events prompted them to reflect on what occurred, their actions, encountered constraints, and potential alternative approaches if given another opportunity to complete the homework.

We found that three child-caregiver dyads maintained journals documenting the events preceding our events-based sessions. These journals facilitated children's recollection of their recent activities and functioned as \textit{lesson plans}, guiding children on the content they intended to teach us.

Asking the child-caregiver dyads to assess our preliminary assumptions provided new insights into the homework information-searching processes observed across the nine case studies.

\begin{quote}
    R: I learned that Russell likes watching videos on YouTube when he has a school project. Russell likes checking if other children’s projects work before trying it. Is that correct?\par
    Russell: Yes.\par
    Alice: Is it? Are you sure?\par
    Russell: No, I watch videos because my mom asks me to check if my idea is going to work before we spend any money.\par
    \textbf{Assess/Reinforce Learning}
\end{quote}

We initially assumed that Russell would search for and watch videos on YouTube to gain inspiration for his school projects. However, in reality, Alice intended for Russell to watch these videos to learn how to execute the project before actually undertaking it. Russell and Alice corrected our initial assumptions, aiding us in refining our understanding and establishing a more robust connection within the interplay of context, agency, and interactions/transactions. We incorporated these corrections into our notes and integrated the new insights into our analysis.
\subsubsection{Contextual Relationships}
The multistage aspect in EXPLORA contributed to collecting multiple accounts and identifying contextual relationships. We found that this approach helped us identify what scenarios dyads deemed ideal or less than ideal for homework information searching.

\begin{quote}
    Philip: Mom, you’re not teaching R correctly. I don’t use the tablet all the time because it’s slow.\par
    Brenda: But you’re using it now. So, we need to teach R what we’re doing now. We can explain that to R when you finish your homework.\par
    \textbf{Philip (\,child)\,- Brenda (\,caregiver)\,, Observation}
\end{quote}

In the interview with Philip and Brenda, we learned that Philip only used the family tablet to fact-check homework inquiries using Google Assistant. In the observation, Philip expressed a desire to use the family computer, but his mother encouraged him to use the tablet to familiarize himself with it. Brenda wanted Philip to learn how to use the tablet properly, given that the family computer is not always accessible.

\begin{figure}[H]
    \centering
    \includegraphics[width=\linewidth,height=10cm,keepaspectratio]{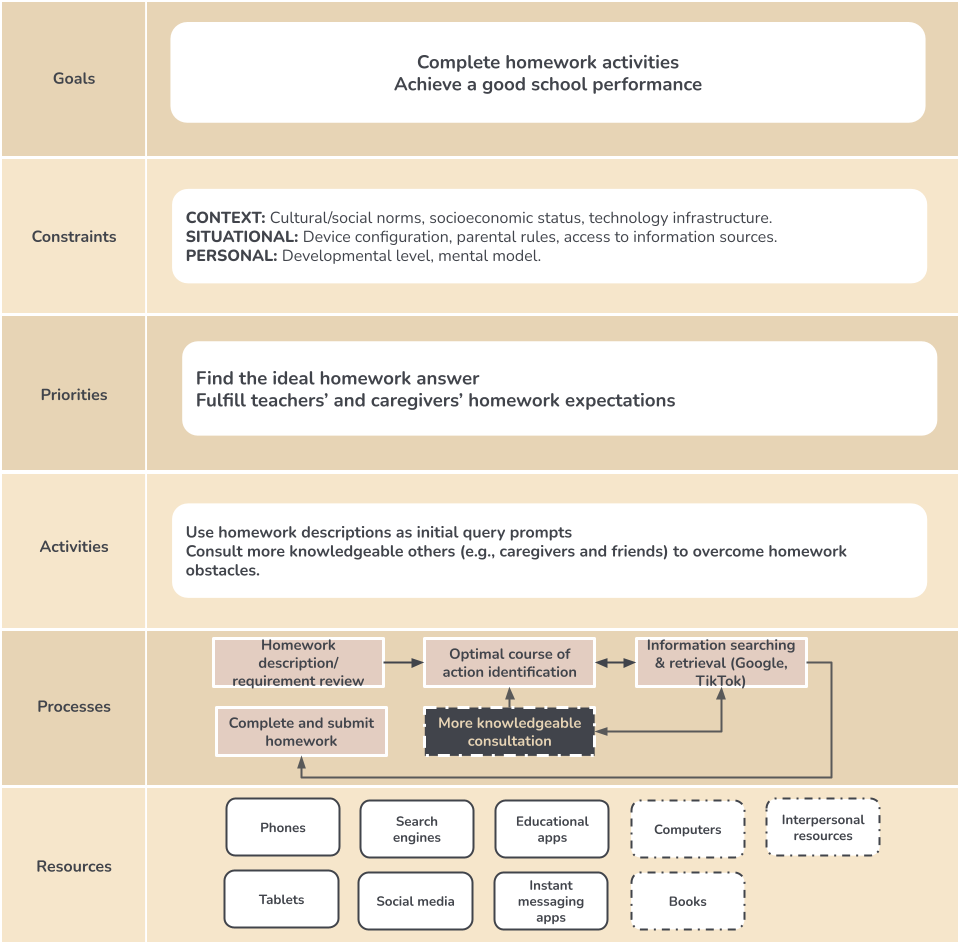}
    \caption{EXPLORA means-ends analysis. The solid-line rectangles represent expected actions, while the dashed-line rectangles represent optional actions.}
    \Description{A template depicting a means-ends analysis, which explores the relationship between context, agency, interactions and transactions. The means-ends analysis focuses on understanding what, how and why involved in human-computer interaction.}
    \label{fig:EXPLORA_MEA}
\end{figure}

We found that the MEA levels enabled us to determine the relevant context and its contextual aspects for the case studies under analysis. Instead of addressing every aspect, our focus was on the contextual elements that posed a constraint (\,e.g., opportunity or limitation)\, for the reported or demonstrated homework information-searching process. For instance, although frequently mentioned by caregivers, we did not deem school policies (\,e.g., homework formatting)\, a critical constraint, as it did not pose a constraint to homework information-searching processes reported, observed and recreated across the nine case studies (\,figure \ref{fig:EXPLORA_MEA})\,.

We began the MEA by identifying the physical resources children frequently used for homework (\,\textit{what})\,. Our focus was on understanding why these specific resources were chosen, influenced by resource availability and knowledge of searching for the homework information needed. Following this, children selected the most suitable homework information-searching strategy (\,\textit{how})\, by reviewing homework descriptions, considering factors like convenience and prior strategies, conducting information-searching and retrieval, consulting more knowledgeable individuals (\,e.g., caregivers, teachers and peers)\, if needed, and preparing and using the information. The analysis delved into how and why children adhered to this specific homework information-searching process, revealing a prioritization of finding the ideal solution to meet academic expectations, balancing constraints and strategies to excel academically.
\subsubsection{Design Recommendations}
We established the relationship between common problems and the strategies children employed to overcome them. For instance, children encountered three critical barriers in their homework information-searching: (\,1)\, ineffective search queries, (\,2)\, difficulty understanding homework problems, and (\,3)\, limited prior knowledge of homework topics. To address these issues, children could either search online or consult more knowledgeable individuals when facing challenges with homework topics. While searching online seemed easier, children needed to evaluate the reliability of information and summarize concepts, especially when caregivers were unavailable. Consequently, children often sought assistance from more knowledgeable individuals, who provided summaries or simplifications of concepts. Engaging with such individuals not only aided in critical thinking but also enhanced their skills in homework information-searching.

\begin{figure}[H]
    \centering
    \includegraphics[width=\linewidth]{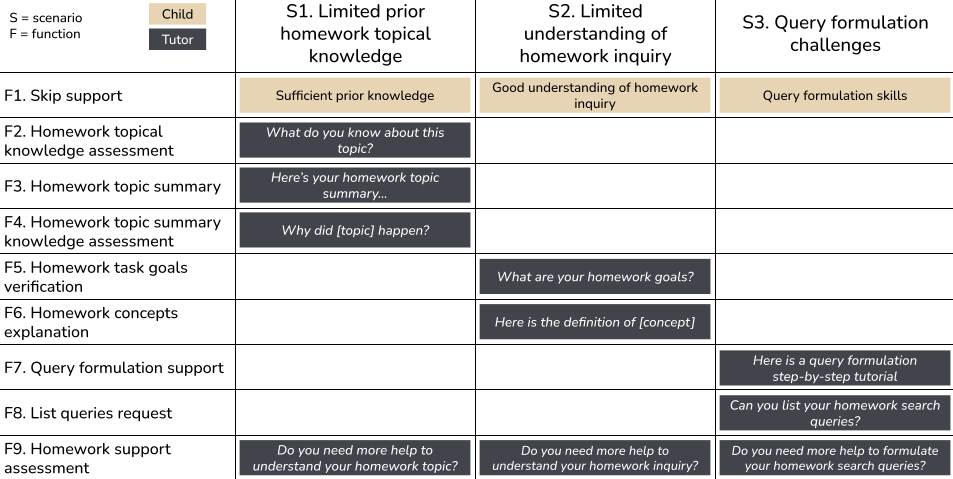}
    \caption{Design Recommendations for a Homework Tutor System. The EXPLORA methodology uncovered common scenarios, enabling the development of recommendations to enhance support for homework information-searching in the analyzed contexts.}
    \Description{A template depicting the frequent scenarios and ideal functions to address the constraints elicited in the nince case studies presented in the paper}
    \label{fig:homeworktutor}
\end{figure}

The common problems and workarounds evolved into scenarios that depict instances where children needed external assistance to decipher problems and navigate constraints. These scenarios encompass (\,1)\, formulating queries for information searching, (\,2)\, framing homework problems, and (\,3)\, acquiring prior knowledge on homework topics. To demonstrate the practicality and applicability of EXPLORA in aiding design recommendations, we provide a condensed version of the \textit{homework tutor} (\,figure \ref{fig:homeworktutor})\,.

Given the frequent scenarios children encountered during our case studies, we formulated a design recommendation for a \textit{homework tutor}. Figure \ref{fig:homeworktutor} illustrates how the functions for the \textit{homework tutor} support each of the three situations children encountered in the Brazilian contexts under analysis. In our design recommendation, we acknowledge the diverse cognitive abilities, prior knowledge, homework information-searching skills, technology access, and agency among children.
\section{Discussion}
The EXPLORA methodology integrates attitudinal and behavioural methods to holistically grasp children's context, interactions, and agency. Employing a multimodal, multistage study design enhances our understanding of children's interactions with technology within various contexts \cite{bonsignore_sharing_2013, cumbo_exploring_2021, marcu_breakdowns_2019, nansen_children_2012}. By aligning our protocols with children's terminology and establishing connections through a \textit{teacher-apprentice} relationship, we created an environment that encouraged children to express their thoughts, demonstrate their homework information-searching processes, and connect them to problem-solving and decision-making strategies relevant to homework information searching.

The EXPLORA methodology offers four core contributions (\,1)\, a structured approach to enhance the identification of relevant interactions and constraints, (\,2)\, a direction towards an in-depth analysis of behavioural data, (\,3)\, a procedure to guide children’s report and demonstration of interactions with technology, and (\,4)\, an approach for prioritizing and formulating design recommendations in CCI studies. We will describe these contributions in detail in the next subsections.
\subsection{A structured approach}
A strategic approach to the method sequence (\,interview-observation-events-based)\, enabled us to capture valuable insights, guiding our focus on the contextual aspects that required further analysis, thus enhancing our understanding in this area \cite{ferron_walk_2019, gillen_day_2018, nansen_children_2012, holtzblatt_contextual_2014}. This finding can assist researchers not only in designing EXPLORA studies but also in conducting contextual inquiries, especially when aiming to understand children's terminologies, frequent interactions, and perceptions before engaging in observations.

While acknowledging potential limitations in generalizing findings, capturing interactions in context offers insights into the constraints children face while engaging with technology. We found that our MEA approach can elicit hypotheses testable in larger experimental datasets. The EXPLORA methodology anticipated contextual constraints on children's homework information-searching processes online, unveiling frequent homework information-searching processes and associated constraints across nine case studies. This addresses a crucial limitation in field studies—identifying relevant interactions and contextual aspects \cite{dourish_implications_2006, shapiro_limits_1994}.

Extensive field engagement holds a pivotal role in cultivating rapport and trust, influencing the depth and richness of participants' narratives \cite{ferron_walk_2019, gillen_international_2010, kalinowski_ecological_2021}. Our study encompassed seven sessions with child-caregiver dyads. While these dyads progressively shared their experiences from the third week onward, we were able to capture rich examples as our study encompassed seven weeks. Moreover, despite the relatively modest sample size, data saturation was attained with child-caregiver dyads, partly attributed to the guiding question (\,"\textit{"Can you teach me [this]?"}) and the means-analysis framework we employed.
\subsection{In-depth analysis of behavioural data}
Previous work has explored the practicalities involving the analysis of field studies data \cite{shapiro_limits_1994}. While these analyses provide detailed descriptions of participants’ interactions in context, it may be challenging to translate these data into practical design recommendations \cite{shapiro_limits_1994, suchman_human-machine_2006}.EXPLORA, through the integration of sequential data collection procedures with MEA, offers a framework for comprehending the dynamics among context, interactions, transactions, and agency.
\subsection{Guidance to children's report and demonstration}
Finding a balance between offering appropriate guidance and preserving natural interactions in context-enriched CCI studies has posed a longstanding challenge \cite{marhan_review_2012}. Although introducing minimal guidance can affect the naturalistic aspect of the study, a more explicit prompt helps children understand expectations, enabling them to provide detailed demonstrations and reports of past experiences when guided \cite{bruckman_hci_2007, cameron_elicitation_1996}. With the \textit{Can you teach me [this]?} prompt, we observed minimal impact on guidance, as the dyads, particularly the children, delivered detailed reports and demonstrations of contextual constraints, interactions, transactions, and agency involving homework information-searching. Unlike \textit{teach-me-back} approaches that use drawings to prompt children to discuss their experiences \cite{ferron_walk_2019, marhan_review_2012}, EXPLORA encourages children to report and demonstrate their homework information-searching processes through interviews, observations and events-based sessions.
\subsection{Prioritizing and formulating design recommendations}
Using MEA, we could anticipate scenarios in which children might encounter challenges in homework information-searching processes. This anticipation helps identify which design recommendations to prioritize. Furthermore, the EXPLORA methodology establishes a foundation for design recommendations, providing a rationale and a pathway for achieving them. This addresses a key concern in HCI regarding how field studies can establish a foundational connection with interaction design \cite{dourish_implications_2006}.
\subsection{Limitations and Future Work}
We encountered several limitations in the EXPLORA methodology. Despite our relatively small sample size (\,nine children and nine caregivers)\,, the EXPLORA methodology yielded valuable insights into homework information-searching processes involving the use of devices and apps in Brazilian contexts. In addition to the case studies presented here, we employed the EXPLORA methodology in an online study with K-12 Brazilian teachers and have plans to apply it in future CCI studies.

While online data collection allowed us to reach Brazilian child-caregiver dyads, we had limitations in capturing background activities and unframed interactions. Despite these limitations, the study's design, marked by its multimodal and multistage structure spanning seven weeks, aimed to minimize the potential loss of contextual information. Given that our study was conducted entirely online, a future iteration could consider incorporating in-person methods to provide a more comprehensive understanding of children's context-specific activities. Another iteration involves integrating context-enrichment surveys allowing participants to answer questions about their perspectives, terminologies and understanding of context, CCI and agency.

A privacy concern that emerged concerned capturing background and bystander data. While the dyads consented that background and bystander data could be captured in our study, we consistently reminded the dyads at the beginning of each session that this could happen. Moreover, we informed the dyads that we would follow the same anonymization process for background and bystander data.

Participant validation guarantees transparency, empowers participants, enhances research quality, and ensures adherence to ethical standards \cite{cameron_day_2018, gillen_day_2018, kalinowski_ecological_2021}. The child-caregiver dyads had the opportunity to assess our assumptions (\,providing veridical member-checking information)\,, aligning them with their intended meanings. While acknowledging the benefits, it is essential to recognize potential limitations, such as participants' hesitancy to provide critical feedback to adults or grappling with intricate concepts. Though some hesitancy was observed in our study, the consistent emphasis on the \textit{teacher-apprentice} relationship mitigated this impact.

Incorporating a participatory design stage can help children visually articulate potential solutions for challenges specific to their context during tasks. In future studies, we aim to introduce a participatory design stage that encourages participants to suggest methods that adeptly capture their interactions within the given context. 
\section{Conclusion}
Gaining insights into children's interactions with technology directly from their natural behaviours and accounts is crucial for designing technologies that align with their daily routines and cognitive processes. The EXPLORA methodology introduces a framework encompassing multiple stages and methods, including attitudinal and behavioural data collection, and means-ends analysis. This approach captures natural child-computer interactions while simultaneously empowering children as teachers of their own experiences with technology. Although demonstrated through single case studies, EXPLORA holds promise for broader investigations that centre on understanding contexts as interconnected systems, establishing a foundational benchmark applicable across diverse settings. By outlining a systematic research process, our objective is to assist CCI researchers in navigating ethical considerations essential for establishing trustworthy relationships with children, caregivers, and educators.

\begin{acks}
We extend our heartfelt gratitude to the children and caregivers who graciously dedicated their time and insights to participate in our study. Without their enthusiastic collaboration, this study would not have been possible. Additionally, we express our gratitude to the anonymous reviewers for their dedicated efforts and insightful feedback during the review process.
\end{acks}

\bibliographystyle{ACM-Reference-Format}
\bibliography{EXPLORA}

\end{document}